\documentclass[a4paper, preprint,aps ]{revtex4}
\usepackage{graphicx}
\usepackage{color}
\usepackage{lscape}
\usepackage{amsmath}
\usepackage{amssymb}

\begin{document}

\title{Block Density Matrix Renormalization Group with Effective Interactions}

\author{Haibo Ma}
\email [Electronic address:~]{haiboma@physik.rwth-aachen.de}
\affiliation{Institut f\"{u}r Theoretische Physik C, RWTH Aachen University, D-52056 Aachen, Germany}
\author{Chungen Liu}
\email [Electronic address:~]{cgliu@netra.nju.edu.cn}
\author{Yuansheng Jiang}\affiliation{
            Institute of Theoretical and Computational Chemistry,
            Nanjing University,
            Nanjing, 210093,
            China }
\date{Latest revised on \today}

\begin{abstract}
Based on the contractor renormalization group (CORE) method and the
density matrix renormalization group (DMRG) method, a new computational
scheme, which is called the block density matrix renormalization
group with effective interactions (BDMRG-EI), is proposed to deal
with the numerical computation of quantum correlated systems. Different from the convential CORE method in the ways of calculating the blocks and the fragments, where the DMRG method instead
of the exact diagonalization is employed in BDMRG-EI, DMRG-EI makes the calculations of larger blocks
and fragments applicable. Integrating DMRG's advantage of high accuracy and
CORE's advantage of low computational costs, BDMRG-EI can be widely
used for the theoretical calculations of the ground state and low-lying excited states of large
systems with simple or complicated connectivity. Test calculations on a 240 site one-dimensional chain and a double-layer
polyacene oligomer containing 48 hexagons demonstrate the efficiency and
potentiality of the method.
\end{abstract}
\maketitle

\section{Introduction}
\label{sec1}The real space renormalization group (RSRG) method \cite{Wilson75},
firstly proposed by Wilson in 1975, is a variational scheme that can
solve large systems through truncating the Hilbert space
successively. Despite RSRG's great success in weakly correlated
systems, lots of theoretical calculations have found that it failed
for strongly correlated systems.\cite{Chui78,Bary79,Hirsh80,Lee79} The breakdown of the RSRG method in the strongly correlated
systems is ascribed to its lack of account of the interactions between
the blocks, i.e. merely considering isolated blocks in RSRG imposes
wrong fixed boundary conditions for the blocks which should actually
be in open boundary conditions.\cite{White92_0}

In order to improve RSRG's performance for strongly correlated
systems, various new schemes have been proposed. These new schemes
can be sorted into two types according to their theoretical philosophies. One
type focused on searching for new optimal criterion of truncation
other than the energy criterion used in the traditional RSRG
methods. Among them, the density matrix renormalization group (DMRG)
method proposed by White in 1992 \cite{White92,White93} has been shown as an
extremely accurate technique in solving one-dimensional (1D)
strongly correlated systems.\cite{Schollwock05} The DMRG method projects the wavefunction of a larger block (the
superblock) onto the system block and then uses the eigenvalue of
the systems block's reduced density matrix as the basis truncation criterion.

The other type focused on taking into account the influence of
discarded excited states in the ``isolated'' block as an account of
the interactions between the blocks, while the energy criterion of
truncation is retained.\cite{Lepetit93,Zivkovic90,Morningstar94,Morningstar96,Weinstein01,Capponi04,Capponi06,Malrieu01,Hajj04,Hajj04_2} In order to achieve this goal, Lepetit and
Manousakis suggested the use of a second-order quasi-degenerate
perturbation theory \cite{Lepetit93}, and Zivkovi\'{c}, et al
suggested defining a new transformed Hamiltonian from the
calculations on dimers of blocks \cite{Zivkovic90}. To formalize the
latter idea in a more general way, Morningstar and Weinstein
introduced the contractor renormalization group (CORE) method in
1994.\cite{Morningstar94,Morningstar96,Weinstein01,Capponi04,Capponi06} CORE introduces effective interactions between the blocks
from the exact spectrum of dimers or trimers of blocks by virtue of
Bloch's effective Hamiltonian theory \cite{Bloch58}. In CORE, the
entire system is divided into blocks with even number of sites and
several eigenstates are kept in each block. In 2001, Malrieu and
Guih\'{e}ry proposed another new improved version of RSRG - real
space renormalization group with effective interactions (RSRG-EI)
\cite{Malrieu01,Hajj04,Hajj04_2}. CORE and
RSRG-EI originated from similar basic ideas, but RSRG-EI uses larger
blocks with odd number of sites and keeps only one eigenstate in
each block and RSRG-EI can be applied to infinite systems due to the
fact that it is iterative. Because CORE and RSRG-EI divide the whole
system into a few blocks and only calculate blocks and fragments
with a certain size exactly instead of the whole system, the
increase of computational costs is only linearly scaling with the size
of the whole system.

Generally, despite of DMRG's great success in 1D systems, the degeneracy in the reduced
density matrix spectrum will increase when DMRG is applied to the quasi-1D large systems with complicated connectivity or
two-dimensional (2D) systems, which implies the impractical
requirement of much more states to be retained to sustain the
numerical precision. In the meantime, although
CORE and RSRG-EI are computationally efficient for large systems,
the accuracy is a bottleneck due to the limitation in the fragment
size that can be solved exactly. Meanwhile, exact or numerically
high-precision calculation of larger blocks and fragments is
actually very essential to achieve reliable account of the intra-block
and inter-block correlation effects, and consequently a reliable
account of the total system. In this paper, we propose a new scheme
- block density matrix renormalization group with effective
interactions (BDMRG-EI), which is a generalization of CORE with the
incorporation of DMRG. It combines the advantage of DMRG in accuracy
and that of CORE in computational costs.

The outline of the paper is as follows: in Section 2, the details of BDMRG-EI methodology are introduced; in
Section 3, demonstrative computations on a 1D chain and a
double-layer polyacene oligomers are presented; finally, we conclude and summarize our results in Section 4.

\section{Methodology}
\label{sec2}
\subsection{Bloch's effective Hamiltonian theory}
The one-to-one correspondence between two isodimensional subspaces
can be well described by the theory of effective Hamiltonians
established by Bloch. Firstly, one may consider a $m$-dimensional
model space, onto which one would like to build an effective
Hamiltonian $\widehat{H}^{eff}$. Let us call the model space as
$S_0$ and its projector as $\widehat{P}_0$
  \begin{equation}\label{ms}
\widehat{P}_0=\sum_{I\in S_0}|I\rangle\langle I|\qquad I=1, \ldots m
  \end{equation}
The main task of $\widehat{H}^{eff}$ is to fulfill the requirement
that its $m$ eigenvalues are exact eigenvalues and its eigenvectors
are the projections of exact eigenvectors in $S_0$. This means that
if
\begin{equation}\label{hp}
\widehat{H}|\psi_k\rangle=\varepsilon_k|\psi_k\rangle\qquad k=1,
\ldots n (n\geq m)
\end{equation}
then $\widehat{H}^{eff}$ satisfies the condition that
\begin{equation}\label{hpe}
\widehat{H}^{eff}\widehat{P}_0|\psi_k\rangle=\varepsilon_k\widehat{P}_0|\psi_k\rangle\qquad
k=1, \ldots m
\end{equation}
The $m$ eigenstates $|\psi_k\rangle$ which are targeted by the
effective Hamiltonian span the so-called target space $S$, with the
projector $\widehat{P}$, isodimensional to $S_0$. If the transform
operator from $S_0$ to $S$ is $\widehat{\Omega}$
($\widehat{P}=\widehat{\Omega}\widehat{P}_0$), then the effective
Hamiltonian satisfies Eq. (\ref{hpe}) is
\begin{equation}\label{he}
\widehat{H}^{eff}=\widehat{P}_0\widehat{H}\widehat{P}=\widehat{P}_0\widehat{\Omega}\widehat{H}\widehat{\Omega}\widehat{P}_0
\end{equation}

On principle, one may choose $S$ arbitrarily provide that
$\widehat{\Omega}$ exists, but an obvious rational choice is to
select $m$ eigenstates with largest projections in $S_0$ to span
$S$, i.e., one should maximize $\sum_{k\in
S}\parallel|\widehat{P}_0\psi_k\rangle\parallel$. After $S$ is
constructed, $\widehat{H}^{eff}$ can be obtained from the following
equation.
\begin{equation}\label{hee}
\widehat{H}^{eff}=\sum_k
|\widehat{P}_0\psi_k\rangle\varepsilon_k\langle\widehat{P}_0\psi_k|
\end{equation}
It should be mentioned that, $\widehat{H}^{eff}$ may be non-Hermitian if $|\widehat{P}_0\psi_k\rangle$ loses
the orthogonality after projection, i.e.,
$\langle\widehat{P}_0\psi_k|\widehat{P}_0\psi_l\rangle\neq\delta_{kl}$. 
Therefore, orthogonalization treatment of
$|\widehat{P}_0\psi_k\rangle$ is necessary before the construction
of $\widehat{H}^{eff}$.
\subsection{DMRG}
Over the last decade, the DMRG method \cite{White92, White93} has
emerged as the most powerful method for the simulation of strongly
correlated 1D quantum systems. Here, we would like to give a brief
introduction to the main physical principles of DMRG. More technique details can be
found in some recent reviews about DMRG \cite{Schollwock05, Schollwock07, Schollwock07_2}.
\subsubsection{Decimation of state spaces}
Due to the exponential growth of the number of degrees of freedom in quantum
many-body systems, the exact simulation in the complete Hilbert
space is apparently not feasible beyond very small sizes. Among a great number of various
approximate simulation methods for quantum many-body systems, one
important class of approximate simulation methods attempts a
systematic choice of a subspace of the complete Hilbert space which
is anticipated to contain the physically most relevant states.
All variational and renormalization group techniques are within this
group, and the essential question is of course to identify the best
decimation strategy which will depend on both the system and the
physical question.

Here, let us particularly focus on the decimation strategy in
strongly correlated 1D quantum systems. Imagine we grow the system
successively, adding site by site. The original system, which we
refer to as an old block, is assumed to be effectively described
within a state space ${|\alpha\rangle}$ of dimension $M$, the new
site within a state space ${|\sigma\rangle}$ of dimension $N$.
Obviously, the state space ${|\beta\rangle}$ of the new block
composed of the old block and the newly added site will have the
dimension $MN$, and for the prevention of exponential growth it will
be decimated down to the dimension $M$. Whatever the physical
decimation strategy one uses, the states of the new block will be a
linear combination of the old states,
\begin{equation}\label{deci}
    |\beta\rangle=\sum_{\alpha}\sum_{\sigma}\langle\alpha\sigma|\beta\rangle|\alpha\rangle|\sigma\rangle\equiv\sum_{\alpha}\sum_{\sigma}A_{\alpha\beta}[\sigma]|\alpha\rangle|\sigma\rangle
\end{equation}
where $N$ matrices $A$ of dimension $M\times M$ have been
introduced, one for each $|\sigma\rangle$, such that the matrix
elements encode the expansion coefficients:
$A_{\alpha\beta}[\sigma]=\langle\alpha\sigma|\beta\rangle$. The introduction of the
$A$-matrices allows to encode the iterative growth of states of larger and larger blocks by matrix multiplications.

Pushing the above idea further it turns out that, we can generally
describe a quantum state of a $L$-site lattice emerging from a
decimation procedure as
\begin{equation}\label{MPS}
    |\psi\rangle=\sum_{\sigma_1...\sigma_L}A^1[\sigma_1]A^2[\sigma_2]...A^L[\sigma_L]|\sigma_1...\sigma_L\rangle
\end{equation} Such states are referred to be matrix product states (MPS) \cite{Klumper91, Fannes92, Klumper92, Derrida93, Ian07} (also known as finitely correlated states).
Now, the main problem for simulating the state of a quantum system effectively
becomes how we can find suitable $A$-matrices with limited
dimensions such that $|\psi\rangle$ approximates the state well. If we want to target the ground state of
a Hamiltonian $\widehat{H}$, apparently the optimal choice for the
$A$-matrices is to find the prescription yielding those $A$ that
minimize $\langle\psi|\widehat{H}|\psi\rangle$ with the constraint of $\langle\psi|\psi\rangle=1$. However, directly
working out this expression leads to a highly non-linear expression
for the energy in $A$, which is numerically very difficult to be
solved. One way to turn the problem low-scaling consists in
providing a starting set of $A$-matrices in a warm-up procedure,
preferably close to the true solution, and then to repeat the
following process iteratively: keeping all $A$-matrices in
$|\psi\rangle$ fixed, with only one or two exception(s), and then
minimizing $\langle\psi|\widehat{H}|\psi\rangle$ with respect to
these one or two flexible $A$-matrix (matrices) and updating these
one or two $A$-matrix (matrices) according to the newly determined
$|\psi\rangle$ with minimized $\langle\psi|\widehat{H}|\psi\rangle$.
After that one shifts the position(s) of the flexible $A$-matrix
(matrices) forth and back through the entire chain successively. An
optimal approximation of the ground state, which is very close to the
real global minimum of $\langle\psi|\widehat{H}|\psi\rangle$, can be
gradually reached. Historically, this process corresponds to the
so-called finite-system algorithm in standard DMRG language.

\subsubsection{DMRG procedures}
However, there are still some questions not answered: how can we derive
$A$-matrices from the newly determined $|\psi\rangle$ and can we
retain all the important information provided by $|\psi\rangle$? Now we will introduce an efficient basis truncation scheme which retains most physically important information and can easily derive new $A$-matrices.

\begin{figure}
\begin{center}
\includegraphics[width =8 cm ]{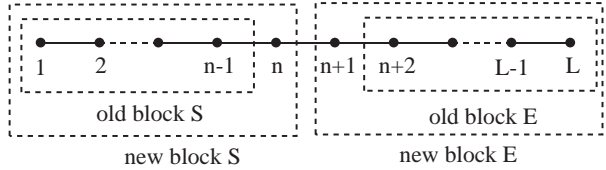}
\caption{\label{fig:DMRG} Block definition in DMRG.}
\end{center}
\end{figure}
Assume we join one block of length $n-1$ and another block with length
$L-n-1$ with two active sites explicitly inserted between these two blocks, as illustrated in Fig.~\ref{fig:DMRG}.
In this case, the left block of length $n-1$ would be described by a
$M$-dimensional Hilbert space with states ${|m^S_{n-1}\rangle}$ which
can be described by $A$ matrices as,
\begin{equation}\label{MPS-2}
    |m^S_{n-1}\rangle=\sum_{\sigma_1...\sigma_{n-1}}A^1[\sigma_1]A^2[\sigma_2]...A^L[\sigma_{n-1}]|\sigma_1...\sigma_{n-1}\rangle
\end{equation}
Similarly the right block would be described by states
${|m^E_{L-n-1}\rangle}$. The left block of length $n-1$ combined with
$n$-th site is now called new block S, and the right block of length
$L-n-1$ combined with ($n+1$)-th site is now called new block E. So the
quantum state of the entire system can be described as,
\begin{equation}\label{fswf}
\begin{split}
    |\psi\rangle&=\sum_{m^S=1}^M\sum_{\sigma^S=1}^N\sum_{\sigma^E=1}^N\sum_{m^E=1}^M\Psi_{m^Sm^E}[\sigma^S\sigma^E]|m^S\sigma^S\rangle|m^E\sigma^E\rangle\\
&=\sum_{i=1}^{N^S}\sum_{j=1}^{N^E}\Psi_{ij}|i\rangle|j\rangle
\end{split}
\end{equation}
where ${|m^S\sigma^S}\equiv{|i\rangle}$,
${|m^E\sigma^E}\equiv{|j\rangle}$ and $N^S=MN$, $N^E=MN$. During the
finite-system DMRG sweeps, we shift the positions of the two active
sites successively to update all $A$ matrices for improving our wavefunction. In order to prevent the
exponential growth of matrix dimension, one has to find a suitable
truncation of the basis ${|i\rangle}$ to $M$ states, whose expansion
in $|m^S\rangle$ and $|\sigma^S\rangle$ will just define the desired
$A^{n}[\sigma_{n}]$. Similarly, we may also have to find a
suitable truncation of the basis ${|j\rangle}$ to $M$ states, whose
expansion in $|m^E\rangle$ and $|\sigma^E\rangle$ will just define
the desired $A^{n+1}[\sigma_{n+1}]$.

In order to find the optimal decimation strategy, one can perform
Schmidt decomposition for
$|\psi\rangle=\sum_{ij}\Psi_{ij}|i\rangle|j\rangle$. After the
Schmidt decomposition, one can get the wavefunction in the following
new form \cite{Schollwock05}:
\begin{equation}\label{SD}
    |\psi\rangle=\sum_{\alpha=1}^{N_{Schmidt}}\sqrt{w_{\alpha}}|w^S_{\alpha}\rangle|w^E_{\alpha}\rangle
\end{equation}
where the scalars $w_{\alpha}$ are non-negative, and
$|w^S_{\alpha}\rangle$ and $|w^E_{\alpha}\rangle$ are newly linear
recombined vectors with $w_{\alpha}$  for S and E parts
respectively. Upon tracing out the E or S part of
the state the reduced density matrix for the S or E part is easily
found to be
\begin{equation}\label{DL}
    \widehat{\rho}_S=\sum_{\alpha=1}^{N_{Schmidt}}w_{\alpha}|w^S_{\alpha}\rangle\langle w^S_{\alpha}|
\end{equation}
\begin{equation}\label{DR}
    \widehat{\rho}_E=\sum_{\alpha=1}^{N_{Schmidt}}w_{\alpha}|w^E_{\alpha}\rangle\langle w^E_{\alpha}|
\end{equation}
Apparently, $w_{\alpha}$ are the eigenvalues of the reduced density
matrix for the S or E part.

The approximate wavefunction where the
space for S or E part has been truncated to be spanned by only $M$
orthonormal states minimizes the distance to $|\psi\rangle$ if one
retains the $M$ eigenstates of $\widehat{\rho}_S$ or
$\widehat{\rho}_E$ with the largest eigenvalues
$w_{\alpha}$.\cite{Schollwock05} This is just the key truncation
criterion in DMRG.

One could also look for a decimation criterion from the
view of maximizing the retained biparticle entanglement between S
and E parts under truncation. As biparticle entanglement is defined
as $S=-\sum_{\alpha}w_{\alpha}log_2w_{\alpha}$ and normally
one has a large number of relatively small eigenvalues, this again
leads to the same truncation prescription: one must retain the $M$
eigenstates of $\widehat{\rho}_S$ or $\widehat{\rho}_E$ with the
largest eigenvalues $w_{\alpha}$. (see e.g. \cite{Schollwock05})

For local quantities, such as energy, magnetization or density, it
was also found that the errors are of the order of the truncation
weight $\epsilon_{\rho}=1-\sum_{\alpha=1}^Mw_{\alpha}$, which
emerges as the key error estimate.

One can therefore derive $A$-matrices from the effective truncation which retains only $M$
eigenstates of with largest eigenvalues $w_{\alpha}$ of the reduced density matrix built by the newly determined $|\psi\rangle$; in DMRG, the most important information for the purpose of minimizing the errors of the approximate wavefunction and maximizing the retained biparticle entanglement between the blocks is kept if one retains the eigenstates of with largest eigenvalues $w_{\alpha}$ of the reduced density matrix.

Obviously, how efficiently DMRG can work depends on how quickly the
ordered eigenvalue spectrum $w_{\alpha}$ of the reduced density
matrix $\widehat{\rho}$ will decay. Empirically, in 1D systems,
density matrix spectra of gapped quantum systems exhibit roughly
system-size independent exponential decay of $w_{\alpha}$. So, DMRG
calculations can efficiently yield reliable accuracy that can be
comparable to exact calculations for 1D quantum systems if one
controls the truncation weight $\epsilon_{\rho}$ very small.

\subsection{Block density matrix renormalization group with effective
interactions}
Similar to that in CORE and RSRG-EI,
in BDMRG-EI, the whole system is also divided into a certain number
($N_{block}$) of blocks with sites and only several low-energy
states are retained in each block. But in BDMRG-EI, we use DMRG
instead of exact diagonization to solve the $m$ lowest eigenstates
for each block. For example, $m$ eigenstates $|\Phi(A)_i\rangle$
with lowest eigen-energies $\epsilon(A)_i$ are obtained for block $A$
with $N_A$ sites. 

At a later stage, we will derive the effective interactions between the neighbor
blocks from the nearly exact spectrum of dimers or trimers of
blocks according to Bloch's effective Hamiltonian theory.

Let me take fragment $A-B$ in the whole system as an example. Fragment $A-B$ contains two blocks: block $A$ and block $B$, and we build small Hilbert spaces
spanned by $m$ lowest eigenstates for each block. Then,
the model space $S_0$ ($dim(S_0)=m^{2}$) is obtained from the direct
product of these two small spaces. The projector of $S_0$ is expressed as $\widehat{P}_0=\sum_{i=1}^{m^2}|\Phi(A-B)_i\rangle\langle\Phi(A-B)_i|$, where $|\Phi(A-B)_i\rangle=|\Phi(A)_j\rangle\otimes|\Phi(B)_k\rangle$.
Then, we perform a standard DMRG calculation for the fragment $A-B$ in oder to get the nearly exact $m^{2}$ low energy eigenstates $|\psi(A-B)_k\rangle$ with the energy $\varepsilon(A-B)_k$ for the block dimer.
Before using Eq. (\ref{hee}) to calculate the the effective Hamiltonian of fragment $A-B$, we should select largest projections
on $S_0$ to build the target space $S$ and perform orthogonalization treatment of $\widehat{P}_0|\psi_k\rangle$. ($\widehat{P}_0|\psi_k\rangle=\sum_i|\Phi_i\rangle\langle\Phi_i|\psi_k\rangle$) 

According to the above introduction to DMRG, $|\Phi(A)_i\rangle$ calculated by DMRG can be 
described as
\begin{equation}\label{MPS}
\begin{split}
|\Phi(A)_i\rangle=&\sum_{\sigma_n}A^{\sigma_1}(i)A^{\sigma_2}(i)\ldots A^{\sigma_{N_A}}(i)\\
&|\sigma_1\rangle\otimes|\sigma_2\rangle\otimes\ldots|\sigma_{N_A}\rangle
\end{split}
\end{equation}
To enforce the boundary conditions, one may require the leftmost matrix $A^{\sigma_1}$ to be $1\times M$-dimensional, and the rightmost matrix $A^{\sigma_{N_A}}$ to be $M\times 1$.
We can write $|\Phi(B)_i\rangle$ and $|\psi(A-B)_i\rangle$  similarly.
\begin{equation}\label{MPS-B}
\begin{split}
|\Phi(B)_i\rangle=&\sum_{\sigma_n}B^{\sigma_{N_A+1}}(i)B^{\sigma_{NA+2}}(i)\ldots A^{\sigma_{N_A+N_B}}(i)\\
&|\sigma_{N_A+1}\rangle\otimes|\sigma_{N_A+2}\rangle\otimes\ldots|\sigma_{N_A+N_B}\rangle
\end{split}
\end{equation}
\begin{equation}\label{MPS-AB}
\begin{split}
|\psi(A-B)_i\rangle=&\sum_{\sigma_n}C^{\sigma_1}(i)C^{\sigma_2}(i)\ldots C^{\sigma_{N_A+N_B}}(i)\\
&|\sigma_1\rangle\otimes|\sigma_2\rangle\otimes\ldots|\sigma_{N_A+N_B}\rangle
\end{split}
\end{equation}
$|\Phi(A-B)_i\rangle$ can be expressed as the direct product of $|\Phi(A)_j\rangle$ and $|\Phi(B)_k\rangle$ as the following equation:
\begin{equation}\label{MPSAB}
\begin{split}
|\Phi(A-B)_i\rangle=&\sum_{\sigma_n}A^{\sigma_1}(j)\ldots A^{\sigma_{N_A}}(j)\otimes B^{\sigma_{N_A+1}}(k)\ldots B^{\sigma_{N_A+N_B}}(k)\\
&|\sigma_1\rangle\otimes|\sigma_2\rangle\otimes\ldots|\sigma_{N_A+N_B}\rangle
\end{split}
\end{equation}
Therefore, the overlap between DMRG states $|\Phi(A-B)\rangle$ and $|\psi(A-B)\rangle$ can be calculated easily according to the following equation:
\begin{equation}\label{Ov}
\begin{split}
\langle\Phi|\psi\rangle&=\sum_{\sigma_n}(A^{\sigma_1*}\ldots A^{\sigma_{N_A}*}\otimes B^{\sigma_{N_A+1}*}\ldots B^{\sigma_{N_A+N_B}*})\\
&\text{  }\text{  }\text{  }\text{  }     (C^{\sigma_1}\ldots C^{\sigma_{N_A+N_B}})\\
&=E_{N_A+N_B}
\end{split}
\end{equation}
where $E_n$ can be calculated successively as the following formula:
  \begin{equation}\label{En}
    E_n=\begin{cases}
    \sum_{\sigma_1}A^{\sigma_1*}\otimes C^{\sigma_1},  &\text{for $n=1$; } \\
    \sum_{\sigma_n}A^{\sigma_n*}E_{n-1}C^{\sigma_n},  &\text{for $1<n\leq N_A$; }\\
    \sum_{\sigma_n}B^{\sigma_n*}\otimes E_{n-1}C^{\sigma_n},  &\text{for $n=N_A+1$; }\\
    \sum_{\sigma_n}B^{\sigma_n*}E_{n-1}C^{\sigma_n},  &\text{for $N_A+1<n\leq N_A+N_B$. }
    \end{cases}
  \end{equation}

When the the overlap between DMRG states can be easily calculated with the above mentioned MPS representations, one can calculate the the effective Hamiltonian of fragment $A-B$ according to Eq. (\ref{hee}). 

Similarly, the effective Hamiltonians of other block dimers, trimers or other fragments can also be derived.

After the determination of the the effective Hamiltonian of different fragments, effective interactions between the blocks can be determined by
subtracting the contributions of all connected sub-clusters as the following equation.

  \begin{equation}\label{heee}
  \begin{split}
&\widehat{H}^{eff}_{ij}=\widehat{H}^{eff}_{i-j}-\widehat{H}^{eff}_{i}-\widehat{H}^{eff}_{j}\\
&\widehat{H}^{eff}_{ijk}=\widehat{H}^{eff}_{i-j-k}-\widehat{H}^{eff}_{i}-\widehat{H}^{eff}_{j}
-\widehat{H}^{eff}_{k}\\
&\text{  }\text{  }\text{  }\text{  }\text{  }\text{  }\text{  }\text{  }\text{  }\text{  }-\widehat{H}^{eff}_{ij}-\widehat{H}^{eff}_{ik}-\widehat{H}^{eff}_{jk}\\
&\ldots
  \end{split}
  \end{equation}

Finally, the effective Hamiltonian for the whole system acting on a truncated
Hilbert space is obtained through the summation of a series of the
local effective Hamiltonians and then directly diagonized with Lanczos or Davidson algorithm.
\begin{equation}\label{tth}
  \widehat{H}^{eff}=\sum_i\widehat{H}^{eff}_{i}+\sum_{ij}\widehat{H}^{eff}_{ij}+
  \sum_{ijk}\widehat{H}^{eff}_{ijk}+\ldots
  \end{equation}

\section{Test Applications}
\label{sec3}
As a test case of our approach, we have implemented BDMRG-EI calculations to a 240
site 1D chain and a double-layer polyacene (Pac(2)) oligomer
containing 48 hexagons within the spin-1/2 Heisenberg model ($
H=J\sum_{\langle ij \rangle}S_iS_j $), where $J$ is a positive
exchange constant, and $S_i$ represents the spin operator of $i$th
site with $\langle ij\rangle$ denoting summation restricted to
nearest neighbors. In our test BDMRG-EI calculations, only two
lowest eigenstates were kept for each block and only two-body
inter-block actions were considered, i.e. $m=2$ and the summation
for the whole system's effective Hamiltonian in Eq. (\ref{tth}) is
restricted to the first two terms.

\subsection{1D chain}
For the 240 site 1D chain, we have four block definition patterns as
follow: I. composing of 10 blocks (24 sites in each block); II.
composing of 6 blocks (40 sites in each block); III. composing of 4
blocks (60 sites in each block); VI. composing of 3 blocks (80 sites
in each block).

\begin{figure}
\begin{center}
\includegraphics[width =9 cm ]{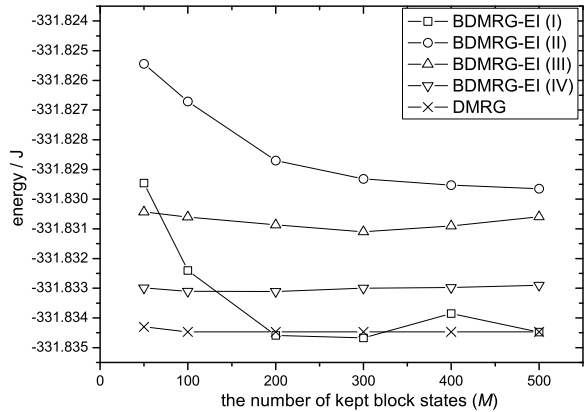}
\caption{\label{fig:fig1} Ground state energy of a 240 site
1D chain.}
\end{center}
\end{figure}

In Fig.~\ref{fig:fig1}, we illustrate the ground state energies
calculated by BDMRG-EI and DMRG with different numbers of kept block
states ($M$). Obviously, direct diagonalization of
the Hamiltonian matrix is impractical at this time for the 240 site
1D Heisenberg chain. Meanwhile, as can be seen in
Fig.~\ref{fig:fig1}, the DMRG calculated energy has converged with
the increase of $M$. Accordingly, this converged value could serve
as a reference value for the comparisons of BDMRG-EI results.

As a whole, calculated energies by BDMRG-EI slow down and converge
gradually with the increase of $M$, although there are some
oscillations, such as the result of BDMRG(I) with $M$=400. This is
due to non-variational nature of the BDMRG-EI method. But
obviously, the range of the oscillation decreases remarkably with
the increase of $M$ and the enlargement of block size.

With large $M$ values, all BDMRG-EI calculations with different
block definitions give satisfactory results in good agreement with
the DMRG value, with largest error within $0.006J$. This implies
that, for 1D systems with simple connectivity, the block size in
BDMRG-EI calculations will not affect the ground state results very
much.

\begin{figure}
\begin{center}
\includegraphics[width =9cm ]{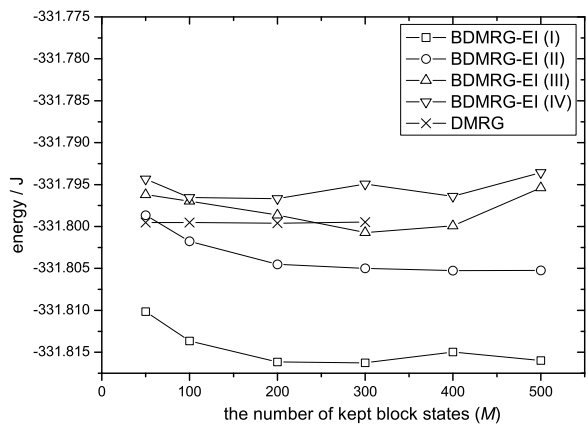}
\caption{\label{fig:fig2} First excited state energy of a
240 site 1D chain.}
\end{center}
\end{figure}

In Fig.~\ref{fig:fig2}, the calculated energies of first excited
state are presented. It can be seen that the errors between the
BDMRG-EI results with DMRG standard values for the excited states
are much larger than those for the ground state. The values of these
errors range from $0.005J$ to around $0.02J$. At the same time, the
influence of block size on the computational accuracy are much more
significant comparing with that in the case of the ground state.
When the block size is too small, BDMRG-EI calculated energies may
incorrectly fall below the DMRG result, and smaller block size will
lead to larger deviation from the DMRG result. This shows BDMRG-EI's non-variational nature again. Only when the block
size is as large as composed of 60 or 80 sites, BDMRG-EI
calculations can give reliable results. These facts elucidated that
electrons are much more delocalized in the excited states, which
makes it more difficult to evaluate the electron correlation effect
in such cases. This implies that incorporation of
DMRG into the CORE method is very necessary in
calculating the excited states of conjugated systems.

\subsection{Double-layer polyacene (Pac(2))}
\begin{figure}
\begin{center}
\includegraphics[width =9 cm ]{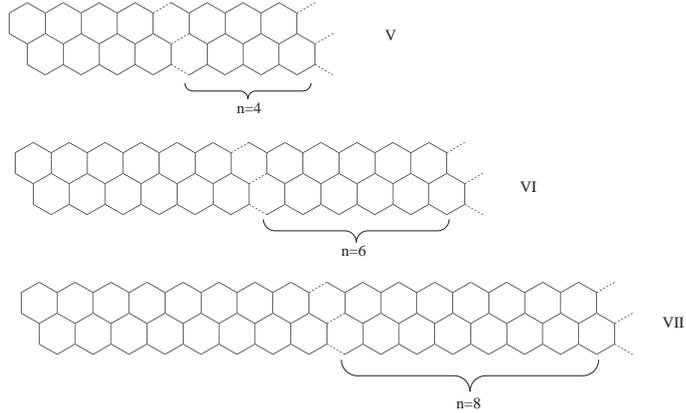}
\caption{\label{fig:fig3} Definitions of blocks for the BDMRG-EI
calculation of a Pac(2) oligomer containing 48 hexagons.}
\end{center}
\end{figure}

The above 1D chain example is a system with simplest connectivity.
When the system's connectivity is more complicated, inter-block
correlations will be more significant. As a consequence, the
influence of block size on the calculation accuracy will be more
remarkable. Here, we take a Pac(2) oligomer containing 48 hexagons
as an example. In our BDMRG-EI calculations, three different block
definitions are adopted, as shown in Fig.~\ref{fig:fig3}.

\begin{figure}
\begin{center}
\includegraphics[width =9 cm ]{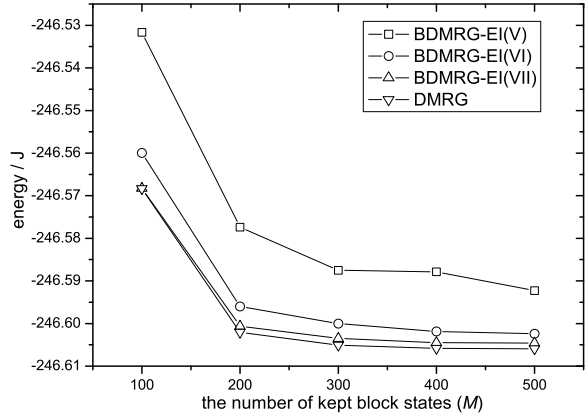}
\caption{\label{fig:fig4} Ground state energy of a Pac(2) oligomer
containing 48 hexagons.}
\end{center}
\end{figure}

\begin{figure}
\begin{center}
\includegraphics[width =9 cm ]{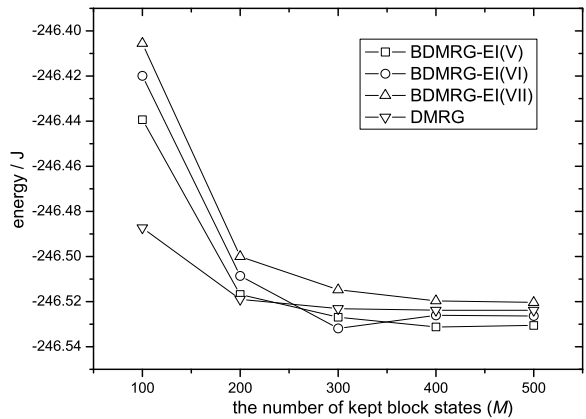}
\caption{\label{fig:fig5} First excited state energy of a Pac(2)
oligomer containing 48 hexagons.}
\end{center}
\end{figure}

Calculated energies of the ground state and the first excited state
of this Pac(2) oligomer are shown in Fig.~\ref{fig:fig4} and Fig.~\ref{fig:fig5}. Similar to
the situation in 1D chain, DMRG calculated values have gradually
converged with the increase of $M$ for this Pac(2) oligomer, and the
DMRG result with $M=500$ serves as the reference value for
comparisons.

In BDMRG-EI calculations with small block sizes, the errors are
around $0.2J$ for both the ground state and the first excited state,
which are definitely intolerable. However, we can still find from
the figures that the results will improve remarkably with the
increase of the block size. As an example, the errors in
BDMRG-EI(VII) for the ground state and the first excited state fall
within $0.05J$. These values are acceptable for such complex system.

It should be mentioned that, even the smallest block size (6-8
hexagons in BDMRG-EI(V)) in BDMRG-EI calculations are far beyond the
applicabilities of the normal CORE or RSRG-EI methods(at most 2-3
hexagons).

\section{Conclusions}
\label{sec4}
This work proposes BDMRG-EI method as an improvement of the traditional CORE. Within the basic
framework of CORE, BDMRG-EI incorporates DMRG calculations into the
solution of the blocks and block dimers or trimers. 

Our demonstrative calculations showed that the block size
effect in systems with complicated connectivity is much more significant
than that in systems with simple connectivity, and that in the
excited states is also much more significant than that in the ground
state. Since the DMRG calculations can achieve high accuracy for systems much larger than
those could be exactly treated, BDMRG-EI could take much more intra-
and inter-block correlations into account comparing to the original CORE and
RSRG-EI methods. Hence, the results of CORE are greatly
improved by BDMRG-EI. Demonstrative calculations of a 240 site 1D
chain and a Pac(2) oligomer containing 48 hexagons convinced that
high accuracy can be achieved in BDMRG-EI calculations when enough large block size is chosen.

Another advantage of BDMRG-EI is its efficiency with low
computational costs. Because it solves only the blocks and fragments
with a certain size by DMRG instead of the whole system, BDMRG-EI
provides a new low-scaling electron structure method which can be
applicable to very large systems with simple or complicated connectivity.

\section*{Acknowledgments}
We are grateful to Ulrich Schollw\"{o}ck, Jean-Paul Malrieu and
Nathalie Guih\'{e}ry for stimulating conversations. This work is
supported by China NSF under the Grant Nos. 20433020 and 20573051.
Ma also thanks the support by Alexander von Humboldt Research
Fellowship.


\begin{thebibliography}{99}

\bibitem{Wilson75}
K. G. Wilson, Rev. Mod. Phys. 47 (1975), 773.

\bibitem{Chui78}
S. T. Chui and J. W. Bary, ibid. 18 (1978), 2426.

\bibitem{Bary79}
S. T. Chui and J. W. Bray, Phys. Rev. B 19 (1979), 4020.

\bibitem{Hirsh80}
J. E. Hirsch, ibid. 22 (1980), 5259.

\bibitem{Lee79}
P. A. Lee, Phys. Rev. Lett. 42 (1979), 1492.

\bibitem{White92_0}
S. R. White and R. M. Noack, Phys. Rev. Lett. 68
(1992), 3487.

\bibitem{White92}
S. R. White, Phys. Rev. Lett. 69 (1992), 2863.

\bibitem{White93}
S. R. White, Phys. Rev. B 48 (1993), 10345.

\bibitem{Schollwock05}
U. Schollw\"{o}ck, Rev. Mod. Phys. 77 (2005), 259.

\bibitem{Lepetit93}
M. B. Lepetit and E. Manousakis, Phys. Rev. B 48
(1993), 1028.

\bibitem{Zivkovic90}
T. P. Zivkovi\'{c}, B. L. Sandleback, T. G. Schmalz, and D. J.
Klein, Phys. Rev. B 41 (1990), 2249.

\bibitem{Morningstar94}
C. J. Morningstar and M. Weinstein, Phys. Rev. Lett. 73 (1994),
1873.

\bibitem{Morningstar96}
C. J. Morningstar and M. Weinstein, Phys. Rev. D 54
(1996), 4131.

\bibitem{Weinstein01}
M. Weinstein, Phys. Rev. B 63 (2001), 174421.

\bibitem{Capponi04}
S. Capponi, A. L\"{a}uchli and M. Mambrini, Phys. Rev. B
70 (2004), 104424.

\bibitem{Capponi06}
S. Capponi, Theor. Chem. Acc. 116 (2006), 524.

\bibitem{Malrieu01}
J. -P. Malrieu and N. Guih\'{e}ry, Phys. Rev. B 63
(2001), 085110.

\bibitem{Hajj04}
M. A. Hajj, N. Guih\'{e}ry, J. -P. Malrieu and B. Bocquillon, Eur.
Phys. J. B 41 (2004), 11.

\bibitem{Hajj04_2}
M. Al Hajj, N. Guih\'{e}ry, J. -P. Malrieu and P. Wind, Phys. Rev. B
70 (2004), 094415.

\bibitem{Bloch58}
C. Bloch, Nucl. Phys. 6 (1958), 329.

\bibitem{Schollwock07}
U.  Schollw\"{o}ck, J. Magn. Mag. Mat. 310 (2007), 1394.

\bibitem{Schollwock07_2}
U.  Schollw\"{o}ck, Int. J. Mod. Phys. B 21 (2007), 2564.
\bibitem{Klumper91}
A. Kl\"{u}mper, A. Schadschneider, and J. Zittartz, J. Phys. A: Math. Gen.
24 (1991), L955.

\bibitem{Fannes92}
M. Fannes, B. Nachtergaele, and R. F. Werner, Commun. Math. Phys. 144 (1992), 443.

\bibitem{Klumper92}
A. Kl\"{u}mper, A. Schadschneider, and J. Zittartz, Z. Phys. B 87 (1992), 281.

\bibitem{Derrida93}
B. Derrida, M. R. Evans, V. Hakim, and V. Pasquier, J. Phys. A: Math. Gen. 26 (1993), 1493.

\bibitem{Ian07}
I. P. McCulloch, J. Stat. Mech. P10014 (2007).

\end{thebibliography}
\end{document}